# Frequency comb generation in a pulse-pumped normal dispersion Kerr mini-resonator

Yiqing Xu,[1, 2] Alexander Sharples,[1, 2] Julien Fatome,[1, 2, 3] Stéphane Coen,[1, 2] Miro Erkintalo,[1, 2] and Stuart G. Murdoch[1, 2, *]

[1] *The Dodd-Walls Centre for Photonic and Quantum Technologies, Auckland 1010, New Zealand*
[2] *Department of Physics, University of Auckland, Auckland 1010, New Zealand*
[3] *Laboratoire Interdisciplinaire Carnot de Bourgogne, UMR6303 CNRS-UBFC, Dijon, France*
*Corresponding author: s.murdoch@auckland.ac.nz

**Kerr microresonators driven in the normal dispersion regime typically require the presence of localized dispersion perturbations, such as those induced by avoided mode crossings, to initiate the formation of optical frequency combs. In this work, we experimentally demonstrate that this requirement can be lifted by driving the resonator with a pulsed pump source. We also show that controlling the desynchronization between the pump repetition rate and the cavity free-spectral range (FSR) provides a simple mechanism to tune the center frequency of the output comb. Using a fiber mini-resonator with a radius of only 6 cm we experimentally present spectrally flat combs with a bandwidth of 3 THz whose center frequency can be tuned by more than 2 THz. By driving the cavity at harmonics of its 0.54 GHz FSR, we are able to generate combs with line spacings selectable between 0.54 and 10.8 GHz. The ability to tune both the center frequency and frequency spacing of the output comb highlights the flexibility of this platform. Additionally, we demonstrate that under conditions of large pump-cavity desynchronization, the same cavity also supports a new form of Raman-assisted anomalous dispersion cavity soliton.**

The ability of passive Kerr resonators to generate coherent frequency combs (Kerr combs) has attracted considerable interest [1,2], with applications ranging from spectroscopy and high-precision metrology [3,4] to LIDAR [5,6] and optical coherence tomography [7]. Key to the generation of highly-coherent combs in Kerr resonators is the ability to excite stable ultra-short temporal structures within the cavity. The most well-known of these structures is the dissipative Kerr cavity soliton (CS) whose existence typically requires a cavity with anomalous dispersion [8,9]. Whilst recent works have shown that the inclusion of higher-order dispersion can also permit the experimental realization of CSs under conditions of weak normal dispersion [10,11], a strong normal dispersion is generally considered prohibitive for CS formation.

Coherent Kerr combs can nonetheless arise in a normally dispersive resonator through interlocking of switching waves (SWs) [11-15]. Such SWs correspond to localized connections between the upper and lower branches of the bistable cavity response and have been shown to enable whole families of temporal structures – both bright and dark – in the normal dispersion regime [14,15]. Interest in the optical frequency combs corresponding to these normal dispersion structures has been high for two reasons. Firstly, many Kerr materials exhibit normal dispersion at wavelengths where one might wish to generate a frequency comb, and secondly, normal dispersion Kerr combs can operate with extremely high conversion efficiencies. Indeed, the highest conversion efficiency Kerr combs reported to date have all used normal dispersion cavities, with conversion efficiencies in excess of 30% reported by several groups [16,20].

A key drawback of normal dispersion SW combs is that they lack a simple means of excitation. This is because the upper branch of a normally dispersive Kerr cavity does not exhibit modulation instability (MI). Typically, to excite normal dispersion combs in microresonators, researchers have been forced to rely on local perturbations to the dispersion of the driven mode family to enable MI, and hence SW excitation. These dispersion perturbations are induced by engineering avoided mode crossings, either between the driven mode family and other resonator modes [16,17], or between the hybridized modes of coupled ring cavities [18-20]. This additional requirement adds significant complication to the design of cavities that support normal dispersion Kerr combs.

In this Letter, we experimentally demonstrate the generation of Kerr combs in a normal dispersion cavity driven by a pulsed pump. Pulsed driving allows isolated SWs to form on both the leading and trailing edges of the intracavity field, with no perturbation to the cavity's dispersion required [13,21]. These SWs then manifest themselves in the spectral domain via the appearance of strong dispersive waves (DWs) [22,23]. We show that, by controlling the desynchronization between the repetition rate of the pulsed driving field and the cavity free-spectral-range (FSR), it is possible to control the spectral positions of the two DWs, providing a simple mechanism to tune the spectral extent of the comb. The dynamics we present here are to be contrasted with the recent experimental report of pulse-driven Kerr combs generated in a weakly-normal cavity with significant third-order dispersion [11]. In particular, in the near-zero-dispersion regime of Ref. [11], comb formation involves a complex interplay between both SWs and solitons that imposes additional restrictions on the range of accessible desynchronizations; our study focuses rather on comb generation via pure SW dynamics in the strong-normal-dispersion regime. Experimentally, we use a novel fiber mini-ring resonator with a radius of just 6 cm to demonstrate a spectrally flat comb with a bandwidth of over 3 THz whose center frequency can be tuned by 2 THz. By harmonically driving this 'mini-resonator' at multiples of its 0.54 GHz FSR, we experimentally demonstrate combs with adjustable frequency spacings between 0.54 and 10.8 GHz [24]. This allows for operation at typical microresonator line spacings in a cavity formed from standard fiber components. In addition, as the comb is generated from SWs that are intrinsically locked to the pulsed driving field [21,25], the frequency spacing of the output comb is automatically locked to the repetition rate of the external drive signal. Finally, our experiments reveal that, under conditions of large pump-cavity desynchronization, the cavity can also support a new type of Raman-assisted CS that exhibits both low-noise and a very flat broadband profile in excess of 14 THz.

We begin our analysis by considering a normal dispersion passive Kerr resonator driven by an ultrashort pulse train (with the repetition rate approximately synchronized to an integer multiple of the cavity FSR, $f_0 \sim m \times \text{FSR}$). In this pulsed-driven regime, SWs can form on the leading and trailing edges of the intracavity field [13,21]. The phase matching condition for the DWs associated with these SWs requires the phase mismatch $\Delta\phi$ accumulated between the pump and each DW over one roundtrip to be zero [22,23]:

$$\Delta\phi = \sum_{k\geq 2} \frac{\beta_k L}{k!}\Delta\omega^k - \Delta T \Delta\omega + (2\gamma PL - \delta_0) = 0, \quad (1)$$

where $\Delta\omega$ is the frequency shift of the DW from the pump frequency, $\beta_k$ are the $k$-th order dispersion coefficients at the pump frequency, $P$ is the peak power of the intracavity pulse, $L$ is the cavity length, $\delta_0$ is the cavity detuning, and $\Delta T = (f_0/m)^{-1} - FSR^{-1}$ is the temporal desynchronization between the cavity roundtrip time and the pump period.

In Fig. 1(a) we plot the phase mismatch calculated from Eq. (1) for three different values of desynchronization, $\Delta T = 32.9, 0$ and $-16.5$ fs. To match the experimental conditions reported below, we consider a 1.8 ps Gaussian drive pulse at 1550 nm, with a peak power of 30 W. The $Q$ of the resonator is set to 6.4 x 10$^7$, with cavity dispersion coefficients $\beta_2 = 7.5$ ps$^2$/km and $\beta_3 = 0.16$ ps$^3$/km, and nonlinear coefficient $\gamma = 2.5$ W$^{-1}$km$^{-1}$. The cavity length is $L = 38$ cm, and the cavity detuning $\delta_0 = 0.2$ rad. The phase mismatches plotted in Fig. 1(a) show clear quadratic profiles for all three desynchronizations, highlighting the dominant role of $\beta_2$ for these parameters. We note that, in contrast to single-pass DW generation [26,27], higher-order dispersion is not required to phasematch DWs in a cavity, as the detuning $\delta_0$ and desynchronization $\Delta T$ provide additional degrees of freedom [22,23]. Figure 1(a) also shows that adjusting the desynchronization $\Delta T$ allows for the positions of the phase-matched DWs to be controlled, suggesting a mechanism to tune the spectral edges of the output comb. This is confirmed in Figs. 1(b-d), where we plot numerically simulated spectra obtained for the three values of $\Delta T$ considered in Fig. 1(a). Our simulations use the well-known Lugiato-Lefever equation (LLE) with the desynchronization modelled by means of a convective drift term [21,28,29] and parameters as quoted above. These simulations show the formation of spectrally flat combs with an extent that can be continuously tuned by controlling the cavity desynchronization. The DW frequencies predicted by Eq. (1), shown as diamonds, circles and squares in Fig. 1(a), are seen to align with spectral edges of each comb, validating this simple phasematching analysis.

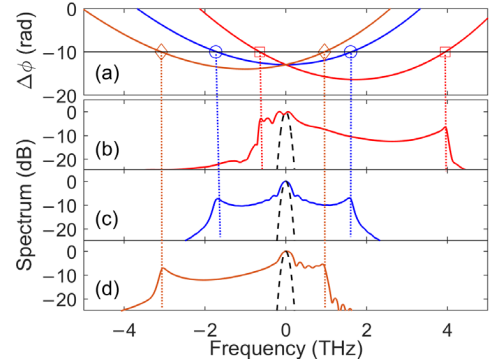

Fig. 1. (a) Phase mismatch as a function of frequency detuning for pump desynchronisations of 32.9 fs (red trace), 0 fs (blue trace) and $-16.5$ fs (orange trace) calculated from Eq. (1). The corresponding phasematched frequencies are marked by diamonds, circles and squares, respectively. (b – d) Numerically simulated spectra for group velocity mismatches of 32.9, 0 and $-16.5$ fs, respectively. The black dashed traces show the input pump spectrum.

To experimentally demonstrate comb generation in a pulse-driven normal-dispersion cavity, we use a fiber mini-ring resonator made from a custom 99/1 fused fiber coupler fabricated directly from a dispersion shifted fiber with a normal dispersion at the pump wavelength of 1550 nm (Corning MetroCor, parameters as quoted above). A single splice is used to form a ring with a radius of only $\sim 6$ cm (38 cm circumference), yielding a cavity with an FSR of 0.54 GHz and a $Q$ of 6.4 x 10$^7$. The pulsed pump is generated from an electro-optic (EO) frequency comb driven by a narrow linewidth CW fiber laser at 1550 nm. A full description of the EO comb source is given in Ref. [24]. The output of the EO comb is amplified and filtered to yield 1.8 ps pulses with an output power of up to 30 W. The exact desynchronization of the pump can be accurately controlled by adjusting the frequency of the RF signal generator used to drive the EO comb. The spectrum at the resonator output is measured by an optical spectral analyzer. In addition, a small fraction of the output is passed through a spectral filter that blocks the input pump frequencies, allowing measurement of the RF spectrum of the comb components generated in the resonator using a 12 GHz photodiode and an electronic spectral analyzer.

We start our experimental investigations with a demonstration of comb generation with the desynchronization set close to zero. Unlike standard anomalous dispersion CSs, the total intracavity power increases monotonically as the pump laser is scanned into resonance from the low wavelength side, with no signatures of 'step-like' features observed [Fig. 2(a)]. This means combs can be excited by directly tuning the drive laser to the desired detuning, at which point they can be maintained through passive thermal locking. In Fig. 2(b) we plot the output comb spectrum observed when the repetition rate of the pump is set close to the FSR of the cavity ($\sim 540$ MHz), as evidenced by the symmetric shape of the spectrum (blue trace). The associated LLE simulation of the output spectrum is overlaid as a dashed trace and shows excellent agreement with the measured spectrum. In Fig. 2(c), we plot the measured fundamental RF beat note of the comb shown in Fig. 2(b): the absence of any excess intensity noise confirms the comb is operating in a stable low-noise regime. Finally, in Figs. 2(d – g) we present the output comb spectra obtained under identical conditions (zero desynchronization), but with the repetition rate of the pulsed drive increased to $f_0 \sim 2$ FSR (1.08 GHz), 6 FSR (3.24 GHz), 10 FSR (5.40 GHz) and 20 FSR (10.8 GHz) respectively. Here, multiple equally spaced pulses circulate simultaneously in the cavity, yielding identical spectral envelopes, but with comb spacings 2, 6, 10 and 20 times the fundamental cavity FSR, respectively. These results show that even in a centimeter-scale ($L = 38$ cm) mini-cavity, GHz line spacings can still be readily achieved through harmonic-driving. Additional measurements (not shown) of the fundamental RF beat notes of these harmonically driven combs allow us to verify that they too are operating in a low-noise regime.

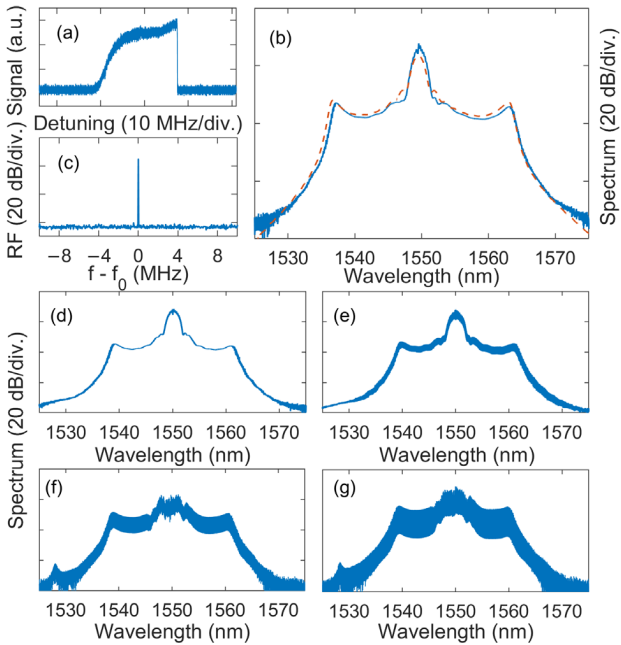

Fig. 2 (a) Nonlinear comb signal measured as the pump is scanned across the resonance. (b) Measured (solid trace) and numerically simulated (dashed trace) output spectrum of the cavity when driven at $f_0 \sim 1$ FSR (540 MHz) and $\Delta T \sim 0$. (c) Measured RF beat note (resolution bandwidth 100 Hz). (d – g) Output spectra with cavity harmonically driven at $f_0 \sim 2, 6, 10$ and 20 FSR respectively.

We next demonstrate how the spectral extent of the generated Kerr combs can be controlled by tuning the pump desynchronization. For this demonstration, we set the pump repetition rate to $f_0 \sim 1$ FSR, though note that harmonic driving works equally as well. To begin, we fine tune the frequency of the drive source such that the output spectrum is symmetric. This corresponds to zero desynchronization, and is plotted in Fig. 3(a) revealing a comb bandwidth of $\sim 26$ nm (3.2 THz) centered around 1550 nm. The drive frequency is then reduced by 4.17 kHz ($\Delta T = +14.3$ fs). As shown in Fig. 3(b), the resulting comb is highly asymmetric, with its spectral center of mass shifted to lower wavelengths. Likewise, increasing the drive frequency by 5 kHz from zero desynchronization ($\Delta T = -17$ fs) shifts the comb center of mass to longer wavelengths [see Fig. 3(c)]. In total we are able to tune the comb's central frequency (defined as the center frequency between the two DWs) by 2 THz. Operation at larger values of $\Delta T$ than presented here is possible, though this comes at the expense of a degraded spectral flatness. Superimposed on-top of the experimental spectra shown in Figs. 3(a – c) are the numerically simulated LLE spectra (dashed traces). Once again excellent agreement is seen between experimental and numerical results.

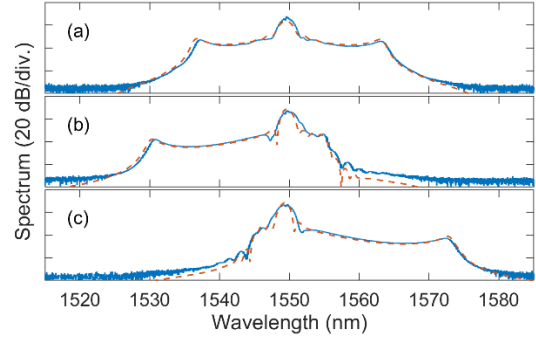

Fig 3. Measured output spectra of the mini-ring resonator for cavity desynchronizations of (a) $\Delta T = 0$ fs, (b) $\Delta T = +14.3$ fs and (c) $\Delta T = -17$ fs (solid traces). The dashed traces show corresponding numerically simulated spectra.

In addition to the tunable combs presented above, we find that the same cavity can also support an unexpected new type of temporal CS when operated under conditions of large temporal desynchronization. The formation of this soliton relies on the strong Raman gain of silica glass, and results in a low-noise spectrally-flat output comb, close to the peak of the Raman gain, at 1650 nm ($\sim 12$ THz below the pump frequency). The weak positive third-order dispersion of the cavity also plays an important role in the soliton's generation, as it sets the dispersion at the soliton frequency to be anomalous [$\beta_2(1650$ nm$) = -4.3$ ps$^2$/km], even though the dispersion at the pump frequency is normal. This allows an anomalous dispersion CS to form at the Raman peak provided sufficient energy can be transferred to this new soliton frequency. The tunable DWs demonstrated in Fig. 3 provide exactly the mechanism required to drive this energy transfer. We note that these formation dynamics require only a single resonator mode family, and differ significantly from previously reported Raman CS observations, whereby a CS in one spatial mode was used to generate a second CS in a different spatial mode directly through stimulated Raman scattering (SRS) [30].

In Fig. 4(a) we plot the phase mismatch given by Eq. (1) for a large negative pump desynchronization of $-45$ fs. Here, Eq. (1) predicts that the long wavelength DW is phasematched at $\sim 1650$ nm, closely coincident with the Raman gain peak. Experimentally, when driving the cavity at this large desynchronization (and $f_0 \sim 1$ FSR) we observe the formation of a broadband frequency comb around 1650 nm [see Fig. 4(b)]. The output comb possesses a very flat broadband spectrum, with its 6 dB bandwidth spanning over 120 nm ($\sim 14$ THz, 26,000 comb lines). In the inset to Fig. 4(b), we plot the measured fundamental RF beat note of the comb. The lack of any excess intensity noise strongly suggests this comb is operating in the low-noise CS regime. This inference is further reinforced through numerical modelling using a generalized LLE equation that includes SRS. The spectral output of this simulation is superimposed on Fig. 4(b) as a dashed trace and shows an excellent agreement with the experimental measurement. In addition, a spectrogram of the simulated intracavity field is plotted in Fig. 4(c). This clearly shows an ultra-short (∼100 fs) localized pulse locked to the leading edge of the intracavity pump pulse which is seen to be strongly dispersed due to the large cavity desynchronization. A detailed zoom of the temporal structure of this ultra-short pulse is plotted Fig. 4(d) and reveals the characteristic temporal oscillations expected of a CS.

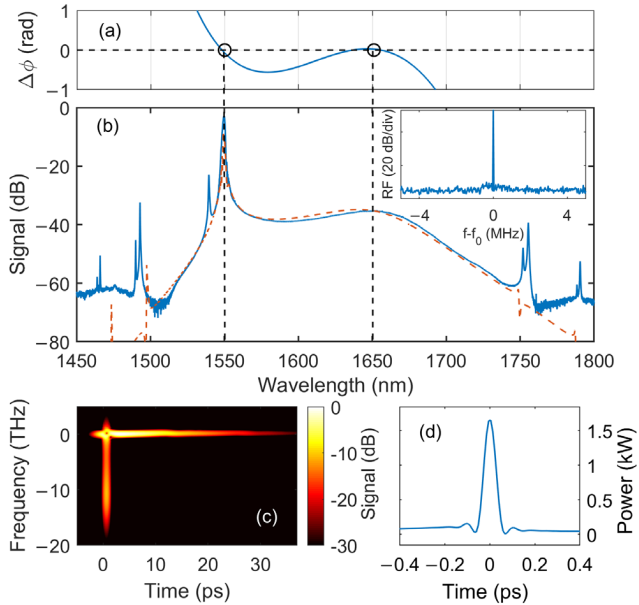

Fig 4. (a) Phase mismatch as a function of wavelength at a desynchronization of −45 fs. (b) Experimentally measured (solid curve) and numerically simulated (dashed curve) Raman-assisted CS spectrum. Inset in (b) shows the measured fundamental RF beat note (resolution bandwidth 100 Hz). (c) Spectrogram of the simulated field, and (d) the corresponding temporal profile of the ultra-short CS pulse.

In conclusion, we have presented a study of comb generation in a pulse-driven normal dispersion Kerr mini-resonator. We show that the spectral extents of the output comb are set by the locations of the two DWs arising from SWs on the leading and trailing edges of the intracavity field, and that these extents can be simply tuned by adjusting the desynchronization. In addition, we are able to drive the cavity at harmonics of its FSR to obtain a tunable comb spacing that is both locked to the external drive signal, and comparable in magnitude to those attainable in microresonators. Furthermore, we are able to demonstrate a new type of Raman-enabled CS in the same cavity. This new soliton yields a broadband spectrally-flat comb detuned by 12 THz from the pump frequency. Finally, we note that the fiber mini-resonator platform we introduce in this Letter could be easily adapted to operate with other fiber types, such as zero-dispersion, dispersion-flattened, birefringent or highly-nonlinear fibers, all of which offer the potential to generate new and interesting forms of Kerr frequency combs.

**Funding.** Marsden Fund and Rutherford Discovery Fellowship of the Royal Society of New Zealand.

**Disclosures.** The authors declare no conflicts of interest.